\documentclass[conference,letterpaper]{IEEEtran}

\usepackage[utf8]{inputenc} 
\usepackage[T1]{fontenc}
\usepackage{url}
\usepackage{ifthen}
\usepackage{cite}
\usepackage{graphicx}
\usepackage{xcolor}
\usepackage{tikz,pgfplots}
\usepackage{comment}
\usetikzlibrary{arrows}

\usepackage[cmex10]{amsmath} 

\usepackage{amssymb}
\usepackage{amsthm}
\newtheorem{theorem}{Theorem}
\newtheorem{lemma}[theorem]{Lemma}

\newtheorem{corollary}{Corollary}

\allowdisplaybreaks

\IEEEoverridecommandlockouts

\begin{document}


\title{On the Capacity of Insertion Channels for Small Insertion Probabilities}

 \author{%
  \IEEEauthorblockN{Busra Tegin and Tolga M. Duman \thanks{This work was funded by the European Union through the ERC Advanced Grant 101054904: TRANCIDS. Views and opinions expressed are, however, those of the authors only and do not necessarily reflect those of the European Union or the European Research Council Executive Agency. Neither the European Union nor the granting authority can be held responsible for them.}}
 \IEEEauthorblockA{Department of Electrical and Electronics Engineering \\
 Bilkent University, Ankara, Turkey\\
                   Email: \tt btegin@ee.bilkent.edu.tr,~duman@ee.bilkent.edu.tr}}





\maketitle 


\begin{abstract}
Channels with synchronization errors, such as deletion and insertion errors, are crucial in DNA storage, data reconstruction, and other applications. These errors introduce memory to the channel, complicating its capacity analysis. This paper analyzes binary insertion channels for small insertion probabilities, identifying dominant terms in the capacity expansion and establishing capacity in this regime. Using Bernoulli$(1/2)$ inputs for achievability and a converse based on the use of stationary and ergodic processes, we demonstrate that capacity closely aligns with achievable rates using independent and identically distributed (i.i.d.) inputs, differing only in higher-order terms.
\end{abstract}


\vspace{-0.65cm}
\section{Introduction}
\vspace{-0.15cm}

Synchronization-error channels, such as deletion, repetition, and insertion channels, effectively model various practical scenarios, including DNA coding \cite{abroshan2019coding, balado2010embedding}, DNA-based storage \cite{heckel2019characterization, lenz2019coding}, magnetic data storage \cite{guan2014coding}, and data reconstruction \cite{levenshtein2001efficient}. Despite their broad range of applications, the capacity of synchronization-error channels remains underexplored. Dobrushin \cite{dobrushin1967shannon} extended Shannon's discrete memoryless channel (DMC) capacity framework \cite{shannon} to channels with memoryless synchronization errors, proving that the capacity for such channels is the limit of the normalized mutual information between the input and the output sequences. Recent works \cite{10619598, morozov2024shannon} consider synchronization-error channels governed by stationary, ergodic Markov chains, generalizing Dobrushin’s theorem and demonstrating equivalence between information and coding capacities.

Despite early results establishing the information stability of memoryless synchronization error channels, determining their exact capacity has proven to be a highly challenging task. In \cite{gallager2000sequential}, Gallager introduced the use of pseudo-random strings in conjunction with convolutional codes and sequential decoding to address synchronization and substitution errors. Various lower bounds have been proposed for different types of synchronization-error channels, such as deletion, insertion, and duplication channels \cite{zigangirov1969sequential,mitzenmacher2006simple, kirsch2009directly}. Upper bounds for channels with synchronization errors have also been extensively studied \cite{diggavi2007capacity,fertonani2010novel, rahmati2013upper, fertonani2010bounds, 10.1145/3281275}. A recent and comprehensive overview of capacity results for synchronization-error channels is available in \cite{cheraghchi2020overview}.


In parallel with efforts to establish capacity bounds, the behavior of the deletion channel in the asymptotic regime of small deletion probabilities has been investigated. For example, \cite{5513746} shows that the trivial lower bound is tight in this limit, providing an upper bound of $1 - (1 - o(1))H(p)$ as $p \to 0$. Relevant to this work, \cite{kanoria2010deletion, kanoria2013optimal} use Dobrushin’s coding theorem \cite{dobrushin1967shannon} to approximate the binary deletion channel capacity with series expansions up to the second and third terms for small $p$. A similar approach is applied to duplication channels in \cite{6457365}, which differ from insertion channels by duplicating transmitted bits rather than inserting random bits, thereby retaining run information. This simplifies the capacity calculation, allowing its computation with any desired accuracy for any duplication probability \cite{4418490}.

In this paper, we study binary insertion channel, where a random bit is inserted after each bit with a given probability. For small insertion probabilities, we analyze the first two terms in the capacity's series expansion, proving their achievability with independent and identically distributed (i.i.d.) Bernoulli$(1/2)$ input and providing a matching converse for the dominant terms. Our main result is as follows: 
\begin{theorem} \label{main_the}
    Let $C(\alpha)$ be the capacity of the insertion channel with insertion probability $\alpha$. Then, for small $\alpha$ and any $\epsilon > 0$, we have
    {\small{
    \begin{align}
        C(\alpha) = 1 + \alpha \log(\alpha)  +G_1\alpha + \mathcal{O}(\alpha^{3/2-\epsilon})
    \end{align}}}
 where 
{\small{
 \begin{align}
     G_1 = -\log(e) &+ \frac{1}{2}\sum_{l=1}^{\infty} 2^{-l-1}l \log l  \nonumber \\
     &+ \frac{1}{2} \sum_{a = 1}^\infty \sum_{b = 1}^\infty  (b+1)2^{-a-b}h\left(\frac{1}{b+1} \right),
 \end{align}}}
which is approximated as $G_1 \approx 0.4901$\footnote{
Specifically, we express \( G_1 \) as
$
G_1 = \Tilde{G}_L 
+ R_L,
$
where 
$
\Tilde{G}_L = -\log(e) + \frac{1}{2} \sum_{l=1}^{L} 2^{-l-1} l \log l 
+ \frac{1}{2} \sum_{a=1}^{L} \sum_{b=1}^{L} (b+1) 2^{-a-b}  h\left( \frac{1}{b+1} \right)
$
and \( R_L \) denotes the summation beyond \( L \), given by
$
R_L \leq \frac{1}{2} \sum_{l=L+1}^{\infty} 2^{-l-1} l^2 
+ \frac{1}{2} \sum_{a=L+1}^{\infty} \sum_{b=1}^{\infty} (b+1) 2^{-a-b} 
+ \frac{1}{2} \sum_{a=1}^{L} \sum_{b=L+1}^{\infty} (b+1) 2^{-a-b},
$
which can be simplified as
$
R_L \leq 2^{-L-2}(L^2 + 8L + 24).
$
In Theorem 1, we approximate \( G_1 \approx \Tilde{G}_L \) for \( L = 1000 \), where the upper bound on \( R_L \) is on the order of \(10^{-296}\), which is negligible compared to \( \Tilde{G}_L \).
}.

\end{theorem}
%

The paper is organized as follows. Section \ref{sec:prelim} introduces the channel model and gives some preliminaries. The maximum rate for the insertion channel is decomposed into entropy terms in Section \ref{sec:IXnYn}. The capacity approximation is provided in Section \ref{sec:firstCh}. The paper is concluded in Section \ref{sec:disc}. In this paper, we exclude detailed proofs of the lemmas and theorems, offering instead discussions on the approaches employed. For details, please refer to the extended version of the paper \cite{tegin2024capacity}.

\textit{Notation:} Throughout the paper, we use base-two logarithms. Additionally, as $\alpha$ approaches 0, we have
$\log (1-\alpha) \approx -\alpha \log(e)$,
where $e$ represents Euler's number.

\section{Preliminaries} \label{sec:prelim}
We consider insertion channel $W_n$ which takes $n$-bit binary vector $X^n$ as input and outputs $Y(X^n)$. The inserted bits are independent of both other insertions and the original input bits. The channel is modeled by random bit insertions after each bit with probability \( \alpha \). We define the insertion pattern by vector \( A^n \in \{0,1\}^n \) whose \( i \)-th element is defined by
\begin{align} \label{ai}
    A_i= 
    \begin{cases}
        1,&\text{if insertion occurs after the }i\text{-th bit of }X^n\\
        0,              & \text{otherwise}
    \end{cases}
\end{align}
for \( i \in \{1,2, \cdots, n\} \).
The corresponding inserted bit vector is denoted by \( B^n  \in \{0,1\}^n \) whose \( i \)-th element is 
\begin{align} \label{bi}
    B_i= 
    \begin{cases}
        0,& \text{if } A_i = 0\\
        U_i,              & \text{if } A_i = 1
    \end{cases}
\end{align}
where \( U_i \) are i.i.d. \(\text{Ber}(1/2) \) random variables. Note that value of $B_i$ for $A_i = 0$ is not important and it is arbitrarily set to 0. 
The expected length of the output vector \( Y(X^n) \) is \( \mathbb{E}[n'] = n + \alpha n \), and we will use the notation \( Y^{n'} \) for the output vector.

Dobrushin in \cite{dobrushin1967shannon} proved the following theorem.
\begin{theorem}  \label{kanoria_theorem} 
(Taken from \cite{dobrushin1967shannon}) For the memoryless channels with finite expectation of output vector length, consider 
    \begin{equation}
    C_{n} \equiv \frac{1}{n} \max_{p_{X^n}} I(X^n; Y(X^n)).
\end{equation}
Then, the following limit exists and defines the capacity of the channels with synchronization errors
\begin{equation}
    C \equiv \lim_{n \rightarrow \infty} C_n = \inf_{n \geq 1} C_n.
\end{equation}
\end{theorem}

Furthermore, as stated in \cite[Th. 5]{dobrushin1967shannon}, the input $(X_1,  \dots, X_n)$ can be considered as consecutive coordinates of a stationary ergodic process as in the following lemma.
\begin{lemma} \label{lem_kanoria}
(Taken from \cite{kanoria2013optimal}) With $X_i \in \{0,1\} $ and stationary and ergodic process $\mathbb X = \{X_i\}_{i \in \mathbb Z}$, the limit $I(\mathbb X) = \lim_{n \rightarrow \infty} \frac{1}{n} I(X^n;Y(X^n))$  exists and we have
 \begin{equation}
    C= \sup _{\mathbb X \in {\mathcal S}} I(\mathbb X),
\end{equation}
where $\mathcal{S}$ denotes  the class of stationary and ergodic processes that take binary values.
\end{lemma}
See \cite{kanoria2013optimal} for the proofs of Theorem \ref{kanoria_theorem} and Lemma \ref{lem_kanoria}.

We define the rate achieved by the stationary and ergodic process \( \mathbb{X} \) as:
\begin{equation} \label{rate_IX}
    I(\mathbb{X}) = \lim_{n \rightarrow \infty} \frac{1}{n} I(X^n;Y(X^n)).
\end{equation}

For the binary process $\mathbb{X}$, the consecutive blocks of 0's and 1's in $X = (X_1, X_2, \cdots, X_n)$ are called runs, and we investigate the capacity of insertion channels with the stationary ergodic input process $\mathbb{X}$. 


%


We analyze the capacity of insertion channel for small insertion probabilities, starting with the perfect channel ($\alpha = 0$) where capacity is 1, achieved using i.i.d. Bernoulli($1/2$) inputs. For small $\alpha$, achievable rates with i.i.d. inputs match the first two terms of the capacity expansion, supported by converse bounds. Using a run-length approach, we limit run lengths and show minimal loss with large thresholds. An upper bound on $I(\mathbb{X})$ for stationary and ergodic processes confirms these rates approximate capacity for small $\alpha$.

\section{Decomposing Rate: Understanding $I(X^n;Y(X^n))$} \label{sec:IXnYn}
In this section, we identify terms to determine the maximum rate for the insertion channel. Input runs are sequences of consecutive bits with the same polarity, and we introduce a vector for output bit lengths (including inserted bits). We use $X(j)$ to represent the $j$-th run in input vector $X^n$, while $Y(j)$ denotes the corresponding sequence of bits in output vector $Y^{n'}$. Let $K$ be the vector of $|Y(j)|$'s, where $|\cdot|$ indicates the number of bits in $Y(j)$. Assuming there are $M$ runs in $X^n$, this can be expressed as:
\begin{align}
    X^{n} &= X(1) \dots X(M-1)X(M), \\
     Y^{n'} &= Y(1) \dots Y(M-1)Y(M), \\
    K &= (|Y(1)|, \dots , |Y(M-1)|, |Y(M)|). \label{def_k}
\end{align} 
We have the following expression for the rate given in \eqref{rate_IX} in terms of entropy using the chain rule:
\begin{align} \label{cap1}
    I(X^{n}; Y) = H(Y) - H(Y,K | X^{n}) + H(K|X^n,Y).
\end{align}
For the term \( H(Y,K | X^{n}) \), we have
\begin{align} \label{yx}
    H(Y ,K| X^{n}) 
    &\stackrel{r_1}{=}  H(Y,K|X^n,A^n,B^n) + H(A^n,B^n|X^n)  \nonumber \\
    & \qquad \qquad \qquad \qquad - H(A^n,B^n|X^n, Y,K) \nonumber  \\
    &\stackrel{r_2}{=}  H(A^n,B^n) - H(A^n,B^n|X^n, Y,K),  
\end{align}
where $ (r_1) $ follows from the entropy chain rule, $ (r_2) $ is due to $ H(Y,K|X^n,A^n,B^n) = 0 $, and independence of $ A^n $ and $ B^n $ from $ X^n $. Consequently, we have:
\begin{align} \label{cap2}
    I(X^{n}; Y) = H(Y) - H(A^n,B^n)  +& H(A^n,B^n|X^n, Y,K) \nonumber \\
    &+ H(K|X^n,Y).
\end{align}
In this expression, \( H(Y) \) measures the randomness of the output, 
\( H(A^n, B^n) \) captures the uncertainty from the insertion pattern and inserted bits, 
\( H(A^n, B^n | X^n, Y, K) \) reflects the residual ambiguity after observing the input, output, and run lengths, 
and \( H(K | X^n, Y) \) quantifies the unpredictability of the run lengths given the input and output.

\section{Capacity Approximation for the Insertion Channel} \label{sec:firstCh}


To establish Theorem \ref{main_the}, the first step involves calculating terms in $I(X^n, Y(X^n))$, as provided in \eqref{cap2}.  We elaborate on these calculations in the subsequent subsections.

\subsection{Calculation of $H(A^n,B^n)$} \label{SecHAB}

We define the process denoting the insertion patterns corresponding to the sequence of insertion channels as $\mathbb{A}$, which is an i.i.d. Bernoulli($\alpha$) process that determines the locations of insertions. Using this process and channel model, we represent the corresponding $n$-bit binary vector as $A^n$, where the $i$-th element is defined in \eqref{ai}, with inserted bit vector $B^n$ in \eqref{bi}.

\begin{lemma} \label{lemma:Hab1}
We have  
  \begin{align}  
  \frac{1}{n} H(A^n, B^n)  = h(\alpha) + \alpha,
\end{align}  
where $h(\cdot)$ is the binary entropy function. 
\end{lemma}
\begin{proof}
Refer to Lemma 4 in the full paper \cite{tegin2024capacity}.
\end{proof}

\subsection{Calculation of $H(A^n, B^n | X^n, Y, K)$} \label{sec:Habxyk}


To calculate $H(A^n, B^n | X^n, Y, K)$, we use an extended run perspective, where each run includes boundary bits of neighboring runs. For a run of 0’s, the last bit of the left and the first bit of the right neighboring run are included; the same applies to runs of 1’s. A modified insertion process, defined by $\hat{A}^n$ and $\hat{B}^n$, assumes at most one insertion per extended run. Multiple insertions are reversed, making input and output runs identical. For small insertion probabilities, typical patterns have sufficiently spaced insertions, contributing to higher-order terms in capacity, allowing accurate estimation of $H(A^n, B^n | X^n, Y, K)$.

%

%
We further define stationary processes $ {\mathbb{Z}}^n$ and $ {\mathbb{V}}^n$ to keep track of reversed insertions as follows: 
\begin{align} 
    {\mathbb{Z}}^n &= \hat{\mathbb{A}}^n \oplus  \mathbb{A}^n,  \label{eqz} \\
    {\mathbb{V}}^n & = \hat{\mathbb{B}}^n \oplus  \mathbb{B}^n. \label{eqv}
\end{align}
With the modified insertion process, if there is no insertion or at most single insertion in an extended run, ${\mathbb{Z}}^n$ and ${\mathbb{V}}^n$ consist entirely of 0's. For the remaining runs, the corresponding segments in ${\mathbb{Z}}^n$ and ${\mathbb{V}}^n$ are the same as in ${\mathbb{A}}^n$ and ${\mathbb{B}}^n$, respectively. Therefore, ${\mathbb{Z}}^n$ and ${\mathbb{V}}^n$ represent the segments in ${\mathbb{A}}^n$ and ${\mathbb{B}}^n$ that are reversed due to the modified process.



In this section, we bound the difference between $H(\hat{A}^n, \hat{B}^n | X^n, \hat{Y}, \hat{K})$ and $H(A^n, B^n | X^n, Y, K)$ for the modified and original insertion processes, where $\hat{Y}$ is the output vector and $\hat{K}$ represents bit lengths including inserted bits with modified process(as in \eqref{def_k}). This bound quantifies the approximation error, which does not affect the first two-order capacity terms. We then compute $H(\hat{A}^n, \hat{B}^n | X^n, \hat{Y}, \hat{K})$, representing ambiguity in the insertion pattern and inserted bits, completing the approximation of $H(A^n, B^n | X^n, Y, K)$.

Consider the following pairs:
\begin{align}
    \big( (X^n, Y, K, A^n, B^n)&, (X^n, \hat Y, \hat{K}, \hat A^n, \hat B^n) \big), \\
    \big( (X^n, Y, K)&, (X^n, \hat Y, \hat{K} ) \big).
\end{align}
Both of these pairs are in the form of $(T,E)$ such that $T$ is a function of $(E, Z^n, V^n)$ and $E$ is a function of $(T, Z^n, V^n)$. 
Using the chain rule and applying some manipulations to these pairs, we can express the following inequality: 
\begin{align} \label{dif_final1}
    |H(A^n, B^n | X^n, Y,K) - H(\hat A^n, \hat B^n &| X^n,  \hat Y, \hat{K})|  \nonumber \\
    &\leq 2H(Z^n,V^n).
\end{align}
For details, we refer to the full version of the paper \cite{tegin2024capacity}.

Clearly, the processes $\mathbb{Z}$ and $\mathbb{V}$ are stationary, and we have $2H(Z^n,V^n)/n \leq 2h(z,v)$. 
To compute the entropy, one can evaluate the PMF of $P(z, v)$, as detailed in \cite{tegin2024capacity}. Using the expected run length $L_0$, the nonzero probabilities in the probability mass function (PMF) can be upper bounded by:
\begin{equation}\label{pzv_bound}
    \begin{aligned}  
	P(z=0,v=0) &\leq 1-2\alpha^2 \mathbb{E}[L_0] + \mathcal{O}(\alpha^3), \\
	P(z=1,v=0) &\leq \alpha^2 \mathbb{E}[L_0] \leq 2\alpha^2 \mathbb{E}[L_0] + \mathcal{O}(\alpha^3), \\
	P(z=1,v=1) &\leq \alpha^2 \mathbb{E}[L_0] \leq 2\alpha^2 \mathbb{E}[L_0] + \mathcal{O}(\alpha^3),
\end{aligned}
\end{equation}
which will be used to calculate the dominant terms of $h(z, v)$. This term identifies the loss incurred when restricting ourselves to the modified insertion process. 

\begin{lemma} \label{lemma:habxyk1}
       Consider any $\mathbb X $ such that  $\mathbb{E}[L_0 \log L_0] < \infty$. Then, 
    \begin{equation}
        \lim_{n \rightarrow \infty } \frac{1}{n}H({A}^n,{B}^n|X^n, {Y}(X^n),K) =  \frac{1}{2}\alpha \mathbb{E} [\log (L_0)] - \eta,
    \end{equation}
    where $ - 2h(z,v) \leq  \eta \leq \alpha^2 \mathbb{E} [L_0 \log (L_0)]+ 2h(z,v)$.
\end{lemma}

\begin{proof}

The proof follows from \eqref{dif_final1} and Lemma \ref{lemmaABXY_marker} below.
\end{proof}

\begin{lemma} \label{lemmaABXY_marker}
    For any $\mathbb X $ such that  $\mathbb{E}[L_0 \log L_0] < \infty$, we have
    \begin{equation}
        \lim_{n \rightarrow \infty } \frac{1}{n}H(\hat{A}^n,\hat{B}^n|X^n, \hat{Y}(X^n),\hat{K}) =  \frac{1}{2}\alpha \mathbb{E} [\log (L_0)] - \delta,
    \end{equation}
    where $ 0\leq  \delta \leq \alpha^2 \mathbb{E} [L_0 \log (L_0)]$.
\end{lemma}

\begin{proof}
The main idea of the proof is as follows. First, we fix the input $x^n$ and any possible output $\hat{y}(x^n)$. Then, we estimate (the $\log$ of) the number of possible realizations of $\hat{A}^n$ and $\hat{B}^n$ that might lead to the pair $(x^n,\hat{y})$ given the marker vector $\hat k$. The final step is to take the expectation over $(x^n,\hat{y},\hat k)$. For details, refer to Lemma 6 in the full paper \cite{tegin2024capacity}.
\end{proof}


\subsection{Calculation of $H(K|X^n,Y)$} \label{SecHKXY}

\begin{lemma} \label{lemma:hkxy_first}
    For a simple insertion channel model with a stationary and ergodic input process $\mathbb{X}$ satisfying $H(\mathbb{X}) > 1 - \alpha^{\gamma}$, where $\gamma > \frac{1}{2}$, we have:
   {\small{
    \begin{align*} 
 \lim_{n \rightarrow \infty} \frac{1}{n} H(K|X^n,Y) =  \frac{\alpha}{2}\sum_{a = 1}^\infty \sum_{b = 1}^\infty  & (b+1)2^{-a} 2^{-b}   h\left(\frac{1}{b+1} \right) + \epsilon_1,
 \end{align*}}}
 where 
   {\small{
 \begin{align}
     - &\alpha^{1+\gamma/2-\epsilon}- 4\alpha^{1+\gamma-\epsilon/2} - \frac{\alpha^2}{2}\sum_{r_j = 1}^\infty \sum_{r_{j+1} = 1}^\infty  (r_j + r_{j+1}) \nonumber \\
     &\qquad \cdot (r_{j+1}+1)h\left(\frac{1}{r_{j+1}+1} \right) \leq \epsilon_1 \leq  4\alpha^{1+\gamma-\epsilon/2} \nonumber \\
     & \qquad \qquad \qquad  \qquad \qquad \qquad \qquad \qquad + \alpha^{1+\gamma/2-\epsilon},
 \end{align}}}
 with $\epsilon> 0$. 
\end{lemma}

\begin{proof}
To compute $H(K|X^n,Y)$, we define a perturbed insertion process $\check{\mathbb{A}}$. For the input process $\mathbb{X}$, we introduce $\check{\mathbb{Z}}$, whose elements correspond to two consecutive input runs with at most one insertion, equated to `0'. Otherwise, they match the original process ${\mathbb{A}}$. To clarify further, for two consecutive input runs $\{S_i, S_{i+1}\}$, we define $\check{\mathbb{Z}}^i$ as a process with all zeros when $\{S_i, S_{i+1}\}$ has at most one insertion. Otherwise, $\check{Z}_l^i = 1$ if $X_l \in S_i$ and $A_l = 1$, $\forall i$. Using $\check{Z}_l^i$, we define $\check{\mathbb{Z}}$ as follows:
\begin{align}
     \check{Z}_l= 
\begin{cases}
    1,& \text{if } \exists i \text{ such that } \check{Z}_l^i = 1\\
    0,              & \text{otherwise.}
\end{cases}
\end{align}
This perturbed process yields a new insertion pattern:
\begin{equation}
    \check{\mathbb{A}} \equiv \mathbb{A} \oplus \check{\mathbb{Z}}.
\end{equation}

%
The output of this perturbed process is denoted by $\check{Y}(X^n)$, and we define $\check{K}$ as follows:
\begin{equation}
      \check{K} = (|\check{Y}(1)|, \dots , |\check{Y}(M-1)|, |\check{Y}(M)|),
\end{equation}
where $\check{Y}(j)$ denotes the bits in the output corresponding to the $j$-th run of the input (including the original bits and insertions due to the perturbed process). We first calculate $H(\check K |X^n,\check Y)$, then bound difference between the original and the perturbed processes. Applying the chain rule, after some intermediate steps, one can write
\begin{align} \label{hkxy1x}
   H&(\check{K}|X^n,\check{Y}) \nonumber \\
   &= \sum_{j=1}^M H\bigg( |\check{Y}(j)|  \Big|  X(j) \cdots X(M),  \check{Y}(j) \cdots \check{Y}(M) \bigg),
\end{align}
where \( X(j) \cdots X(M) \) and \( \check{Y}(j) \cdots \check{Y}(M) \) represent concatenations of \( X(j) \)'s and \( \check{Y}(j) \)'s, respectively. Further details are given in the full paper \cite{tegin2024capacity}.

The proof is completed by using Lemmas \ref{lem:checkK} and \ref{lem_Kdif}.
\end{proof}

\begin{lemma} \label{lem:checkK}
For an insertion channel with the perturbed insertion process and a stationary and ergodic input process $\mathbb{X}$ satisfying $H(\mathbb{X}) > 1 - \alpha^{\gamma}$, where $\gamma > \frac{1}{2}$, we have:
    \begin{align} 
\lim_{n \rightarrow \infty} \frac{1}{n}  H(\check K|X^n, \check Y) =  \frac{\alpha}{2}& \sum_{r_j = 1}^\infty  \sum_{r_{j+1} = 1}^\infty   (r_{j+1}+1)2^{-r_j} 2^{-r_{j+1}}  \nonumber \\
 & \cdot h\left(\frac{1}{r_{j+1}+1} \right) - \epsilon_2 + \epsilon_3,
\end{align}
which is approximately $\approx 1.2885\alpha - \epsilon_2 + \epsilon_3$ where
\begin{align*} \label{eps}
   0 \leq \epsilon_2 \leq  \frac{\alpha^2}{2}\sum_{r_j = 1}^\infty \sum_{r_{j+1} = 1}^\infty & (r_j + r_{j+1}) (r_{j+1}+1) h\left(\frac{1}{r_{j+1}+1} \right),
\end{align*}
\begin{equation}
   - \alpha^{1+\gamma/2-\epsilon} \leq \epsilon_3 \leq \alpha^{1+\gamma/2-\epsilon}.
\end{equation}
\end{lemma}
\begin{proof}
We outline the key steps of the proof here. For further details, please refer to \cite{tegin2024capacity}. The proof involves calculating:
\begin{equation*}
t_j \equiv H\left(|\check{Y}(j)| \,\Big|\, X(j), \dots, X(M), \check{Y}(j), \dots, \check{Y}(M) \right),
\end{equation*}
by analyzing various cases that contribute nonzero values. With a slight change in notation, let us define $\Bar{Y}(j')$ as the runs in $\check Y$, meaning that $\Bar{Y}(j')$ is directly determined by the bits of $\check Y$. The only nonzero contribution comes from the case with $|\Bar{Y}(j)| = |X(j)|$ and $|\Bar{Y}(j+1)| = |X(j+1)|+1$, where we cannot distinguish between two different scenarios. Consider the following example:
    \begin{align}
        X(j)X(j+1) & = [1 \ 1\ 1\ 0\ 0\ 0\ 0] \\
        \Bar{Y}(j)\Bar{Y}(j+1) & = [1 \ 1\ 1\ 0\ 0\ 0\ 0\ 0]
    \end{align}
    Here, we cannot determine whether the `0' bit is inserted at the end of $X(j)$ or inside $X(j+1)$ covered by the two cases:
    \begin{itemize}
        \item Case $\mathcal{V}_1$: If the `0' bit is inserted at the end of $X(j)$ which has a probability:
    \begin{equation}
        P(\mathcal{V}_1) = \frac{1}{2} \alpha (1- \alpha)^{|X(j)|+|X(j+1)|-1}.
    \end{equation}

        \item Case $\mathcal{V}_2$: If the `0' bit is inserted at any of the possible $|X(j+1)|$ locations in $X(j+1)$ which has a probability:
    \begin{equation}
        P(\mathcal{V}_2) = \frac{1}{2} \alpha(1- \alpha)^{|X(j)|+|X(j+1)|-1} |X(j+1)|.
    \end{equation}

    \end{itemize}
Conditioned on this nonzero contribution case and defining $r_j \equiv |X(j)|$, we can normalize the probabilities which gives 
\begin{equation}
    t_j = h\left(\frac{1}{r_{j+1}+1} \right).
\end{equation}
With some intermediate steps, as detailed in the full version \cite{tegin2024capacity}, the desired expected contribution can be obtained as:
\begin{align} \label{hkxy_last}
 H(\check K|X^n,Y) =  A_1 \alpha - \epsilon_2 + \epsilon_3,
\end{align}
where
  {\small{
\begin{align*}
   & A_1 = 
    \frac{1}{2}\sum_{r_j = 1}^\infty \sum_{r_{j+1} = 1}^\infty  (r_{j+1}+1) 2^{-r_j} 2^{-r_{j+1}} h\left( \frac{1}{r_{j+1}+1} \right), \\ 
&   0 \leq \epsilon_2 \leq  \frac{\alpha^2}{2}\sum_{r_j = 1}^\infty \sum_{r_{j+1} = 1}^\infty  (r_j + r_{j+1}) (r_{j+1}+1)  h\left(\frac{1}{r_{j+1}+1} \right),  \label{epsilon2} \\
&   - \alpha^{1+\gamma/2-\epsilon} \leq \epsilon_3 \leq \alpha^{1+\gamma/2-\epsilon}.
\end{align*}}}
Note that we have $A_1 \approx 1.2885$.
\end{proof}

\begin{lemma} \label{lem_Kdif}
With a stationary and ergodic input process $\mathbb{X}$ satisfying $H(\mathbb{X}) > 1 - \alpha^{\gamma}$, where $\gamma > \frac{1}{2}$; the difference between $H(K|X^n,Y)$ and $H(\check{K}|X^n, \check{Y})$ satisfies:
\begin{align}
\lim_{n \rightarrow \infty} \frac{1}{n} |H(K|X^n,Y) - H(\check{K}|X^n, \check{Y})| \leq 4 \alpha^{1+\gamma-\epsilon/2},
\end{align}
\end{lemma}
\begin{proof}
The proof strategy mirrors that of \cite[Lemma V.18]{kanoria2013optimal}. We utilize the perturbed insertion process $\check{Z}$ and define $U(X^n, A^n, Z^n) \in \{t, 0, 1\}^{|{Y}|}$ for each bit in the output of size $|Y|$. Sequentially, for each bit in $Y$, $U$ is defined as
\begin{itemize}
    \item For each bit in both $Y$ and $\check{Y}$, the corresponding element of $U$ is $t$, representing bits originally in $X$ and insertions from the perturbed process.
\item For each bit in $Y$ but not in $\check{Y}$, $U$ is assigned `0' or `1' based on the bit in $Y$, representing reversed insertions from the process difference.
\end{itemize}

Clearly, we have the pairs $  (X^{n}, Y)  \mathrel{\mathop {\longleftrightarrow}\limits ^{U}} (X^{n}, \check {Y})$ and $ (X^{n}, Y, K)  \mathrel{\mathop {\leftarrow \joinrel\relbar\joinrel\relbar\joinrel\rightarrow}\limits ^{(U, Z)}} (X^{n}, \check {Y}, \check {K})$ which results in
\begin{align}
|H(\check{K}(X^{n}) | X^{n}, \check{Y}(X^{n})) - H&({K}(X^{n}) | X^{n}, {Y}(X^{n})) |  \nonumber \\
& \leq 2H(U) + H(Z).
\end{align}

The proof is concluded by demonstrating that
$\lim_{n \rightarrow \infty} \frac{2H(U) + H(\check{Z})}{n} \leq 4\alpha^{1+\gamma-\epsilon/2}$.
The detailed steps are in the full version of the paper \cite{tegin2024capacity}.
\end{proof}

\subsection{Calculations for $H(Y|X)$}
Using the results given in Sections \ref{SecHAB}, \ref{sec:Habxyk} and \ref{SecHKXY}, we obtain the following corollary. 
\begin{corollary} \label{cor1}
For an input process that is stationary and ergodic satisfying $H(\mathbb{X}) > 1 - \alpha^{\gamma}$, where $\gamma > \frac{1}{2}$, we have
  {\small{
\begin{align}
     \lim_{n \rightarrow \infty }  & \frac{1}{n}  H(Y | X^{n}) \nonumber \\
     & = h(\alpha)  + \alpha  \bigg( 1- \frac{1}{2} \mathbb E [\log L_0]  -\frac{1}{2} \sum_{l_j = 1}^\infty \sum_{l_{j+1} = 1}^\infty  (l_{j+1}+1) \nonumber \\
     & \qquad \cdot 2^{-l_j}2^{-l_{j+1}}h\big(\frac{1}{l_{j+1}+1} \big) \bigg) + \zeta,
\end{align}}}
where $ \zeta = \eta + \epsilon_1$ with
  {\small{
\begin{align}
 & - \alpha^{1+\gamma/2-\epsilon} - 4\alpha^{1+\gamma-\epsilon/2} - \frac{\alpha^2}{2}\sum_{r_j = 1}^\infty \sum_{r_{j+1} = 1}^\infty  (r_j + r_{j+1}) \nonumber \\
 & \qquad \qquad \qquad \cdot (r_{j+1}+1) 2^{-r_j}2^{-r_{j+1}}h\left(\frac{1}{r_{j+1}+1} \right) \leq \epsilon_1 \nonumber \\
 & \qquad \qquad \qquad \qquad \qquad \quad \qquad  \leq  4\alpha^{1+\gamma-\epsilon/2} + \alpha^{1+\gamma/2-\epsilon}, \nonumber \\
    &  -2h(z,v) \leq  \eta \leq \alpha^2 \mathbb{E} [L_0 \log (L_0)]+ 2h(z,v), \nonumber
\end{align}}}
with $\epsilon> 0$. 
\end{corollary}
\begin{proof}
    We have
    \begin{align*}
        H(&Y | X^{n}) =  H(Y,K | X^{n}) - H(K|X^n,Y) \\
        &= H(A^n,B^n) - H(A^n,B^n|X^n, Y,K)  - H(K|X^n,Y).
    \end{align*}
    Using Lemmas \ref{lemma:Hab1}, \ref{lemma:habxyk1} and \ref{lemma:hkxy_first} completes the proof. 
\end{proof}

\subsection{Achiveability} \label{sec:ach1}

\begin{lemma}[Achievability]
    Let $\mathbb{X}^*$ be the iid Bernoulli(1/2) process. For any $\epsilon > 0$, we have
     \begin{align}
        C(\alpha) = 1 + \alpha \log(\alpha)  +G_1 \alpha +  \mathcal{O}(\alpha^{3/2-\epsilon}).
    \end{align}
\end{lemma}

\begin{proof}
Since $\mathbb{X}$ is i.i.d. Bernoulli(1/2) with size $n$, it has a run-length distribution $p_L(l) = 2^{-l}$. The corresponding output $Y(X^{*,n})$ is also i.i.d. Bernoulli with length $L = n+\text{Bin}(n,\alpha)$. Hence, shown in the full version \cite{tegin2024capacity}, we have 
\begin{align}
H(Y(X^{*,n})) &= \mathcal{O}(\log(n)) + n(1+\alpha).
\end{align}
Using $H(Y|X^{*,n})$ from the Corollary \ref{cor1} with $P(z, v) = \mathcal{O}(\alpha^2)$ and $\mathbb{E}[L_0 \log L_0 ] < \infty$ completes the proof.
\end{proof}

\subsection{Converse}  \label{sec:conv1}

%
We denote processes with maximum run length $L$ by $\mathcal{S}_L$. Without loss of generality, we restrict ourselves to finite-length runs. Next, we illustrate that restricting ourselves to $\mathcal{S}_{L^*}$ for large enough $L^*$ will not cause a significant loss.

\begin{lemma}
    For any $\epsilon > 0$ there exists $\alpha_0 = \alpha_0(\epsilon) > 0$ such that the following happens for all $\alpha < \alpha_0$. For any $\mathbb X \in \mathcal S $ such that $H(\mathbb X ) > 1 + 2\alpha \log \alpha$ and for any $L^*> \log(1/\alpha)$, there
exists  $\mathbb X_{L^*} \in \mathcal S_{L^*} $ such that
\begin{equation}
    I(\mathbb X) \leq I( \mathbb X_{L^*}) + \alpha^{1/2-\epsilon}(L^*)^{-1}\log(L^*).
\end{equation}
\end{lemma}

\begin{proof}
   We construct $\mathbb{X}_{L^*}$ by flipping a bit each time it is the $(\mathcal{S}_{L^*}+1)$-th consecutive bit with the same value. The density of such bits is upper bounded by $\beta = \frac{P(L_0 > {L^*})}{L^*}$. The corresponding output is denoted by $Y_{L^*} = Y (X^n_{L^*} )$.    
    
    We further define $F = F(\mathbb X, \mathbb A)$ as a binary vector with the same length as the output. The elements of $F$ are 1 wherever the corresponding bit in $Y_{L^*}$ is flipped relative to $Y$, and 0 otherwise. The entropy $H(F)$ can be upper bounded by: 
    %
    \begin{align}
        H(F) \leq h(\beta) (1+\alpha)n + \log(n+1),
    \end{align}
as explained in detail in the full version \cite{tegin2024capacity}. The rest follows the same steps as \cite[Lemma III.2]{kanoria2010deletion}. 
\end{proof}
\begin{lemma}[Converse]
    For any $\epsilon > 0$ there exists there exists $ \alpha_0 =  \alpha_0(\epsilon) > 0$ such that the following happens. For any $L^* \in \mathbb N$ and any $\mathbb X \in \mathcal S_{L^*}$ if $ \alpha <  \alpha_0(\epsilon)$, then
    \begin{equation}
        I(\mathbb X ) \leq 1+  \alpha \log  \alpha+ G_1  \alpha +  \alpha^{2-\epsilon}(1+ \alpha^{1/2}L^*).
    \end{equation}
\end{lemma}
\begin{proof}
We have $H(Y) \leq n(1+ \alpha) + \log(n+1)$,
since $Y(X^n)$ contains $n+\text{Binomial}(n, \alpha)$ bits.
The proof involves using the lower bound in Corollary \ref{cor1} along with $h(z,v)$. Note that for $H(\mathbb{X}) > 1+2\alpha \log \alpha$, it is also true that $H(\mathbb{X}) > 1-\alpha^{1-\epsilon}$. Hence Corollary \ref{cor1} with $\alpha = 1- \epsilon$ applies. 
Furthermore, using the bound on the PMF of $Z$ and $V$ as given in \eqref{pzv_bound} and employing \cite[Lemma IV.3]{kanoria2010deletion}, which gives $\mathbb{E}[L_0] \leq K_1 (1+ \sqrt{\alpha (\log(1/\alpha))^3L^*})$; we obtain
\begin{equation} \label{hzv_bound}
    h(z,v) \leq 0.5\alpha^{2-\epsilon}(2+0.5\alpha^{1/2}L^*),
\end{equation}
for all $\alpha \leq \alpha_0$ with
$\alpha_0 = \alpha_0(\epsilon) > 0$. Also noting $
	\big|\mathbb{E}[\log L_0]- \sum_{l=1}^\infty 2^{-l-1} l \log l \big| = o(\alpha^{1/2-\epsilon} \log L^*)$, the proof follows.
\end{proof}

We further note that the approach used for the insertion channel can also be extended to the Gallager insertion channel, as demonstrated in the full paper \cite{tegin2024capacity}.

\section{Conclusions} \label{sec:disc}


In this paper, we studied the capacity of insertion channels with small insertion probabilities, a highly relevant model for different practical applications including DNA storage. 
Specifically, we identified the dominant terms of the channel capacity for small insertion probabilities and established the capacity in the asymptotic regime.
To derive our results, we used Bernoulli$(1/2)$ inputs for achievability and combined this with a converse argument employing stationary and ergodic processes as inputs. This approach demonstrated that the channel capacity differs from the achievable rates with i.i.d. inputs only in the higher-order terms, providing an accurate approximation in the small insertion probability regime.
For future work, this approach can be extended to nonbinary inputs such as 4-ary alphabets for DNA storage, and to alternative channel models including deletion, insertion-substitution, and absorption channels. These extensions would significantly enhance our understanding of the capacity of channels with synchronization errors.




\bibliographystyle{IEEEtran}
\bibliography{bibs_insertion}

\begin{thebibliography}{10}
\providecommand{\url}[1]{#1}
\csname url@samestyle\endcsname
\providecommand{\newblock}{\relax}
\providecommand{\bibinfo}[2]{#2}
\providecommand{\BIBentrySTDinterwordspacing}{\spaceskip=0pt\relax}
\providecommand{\BIBentryALTinterwordstretchfactor}{4}
\providecommand{\BIBentryALTinterwordspacing}{\spaceskip=\fontdimen2\font plus
\BIBentryALTinterwordstretchfactor\fontdimen3\font minus
  \fontdimen4\font\relax}
\providecommand{\BIBforeignlanguage}[2]{{%
\expandafter\ifx\csname l@#1\endcsname\relax
\typeout{** WARNING: IEEEtran.bst: No hyphenation pattern has been}%
\typeout{** loaded for the language `#1'. Using the pattern for}%
\typeout{** the default language instead.}%
\else
\language=\csname l@#1\endcsname
\fi
#2}}
\providecommand{\BIBdecl}{\relax}
\BIBdecl

\bibitem{abroshan2019coding}
M.~Abroshan, R.~Venkataramanan, L.~Dolecek, and {A. Guillén i Fàbregas},
  ``Coding for deletion channels with multiple traces,'' in \emph{IEEE
  International Symposium on Information Theory (ISIT)}, Paris, France, Jul.
  2019, pp. 1372--1376.

\bibitem{balado2010embedding}
F.~Balado, ``On the embedding capacity of {DNA} strands under substitution,
  insertion, and deletion mutations,'' in \emph{Media Forensics and Security
  II}, vol. 7541.\hskip 1em plus 0.5em minus 0.4em\relax San Jose, CA, USA:
  SPIE, Jan. 2010, pp. 411--422.

\bibitem{heckel2019characterization}
R.~Heckel, G.~Mikutis, and R.~N. Grass, ``A characterization of the {DNA} data
  storage channel,'' \emph{Scientific reports}, vol.~9, no.~1, p. 9663, Nov.
  2019.

\bibitem{lenz2019coding}
A.~Lenz, P.~H. Siegel, A.~Wachter-Zeh, and E.~Yaakobi, ``Coding over sets for
  {DNA} storage,'' \emph{IEEE Transactions on Information Theory}, vol.~66,
  no.~4, pp. 2331--2351, Apr. 2019.

\bibitem{guan2014coding}
Y.~L. Guan, G.~Han, L.~Kong, K.~S. Chan, K.~Cai, and J.~Zheng, ``Coding and
  signal processing for ultra-high density magnetic recording channels,'' in
  \emph{2014 International Conference on Computing, Networking and
  Communications (ICNC)}, Honolulu, HI, USA, Feb. 2014, pp. 194--199.

\bibitem{levenshtein2001efficient}
V.~I. Levenshtein, ``Efficient reconstruction of sequences,'' \emph{IEEE
  Transactions on Information Theory}, vol.~47, no.~1, pp. 2--22, Jan. 2001.

\bibitem{dobrushin1967shannon}
R.~L. Dobrushin, ``Shannon's theorems for channels with synchronization
  errors,'' \emph{Problemy Peredachi Informatsii}, vol.~3, no.~4, pp. 18--36,
  1967.

\bibitem{shannon}
C.~E. Shannon, ``A mathematical theory of communication,'' \emph{The Bell
  System Technical Journal}, vol.~27, no.~3, pp. 379--423, Jul. 1948.

\bibitem{10619598}
R.~Morozov and T.~M. Duman, ``On the capacity of channels with {Markov}
  insertions, deletions and substitutions,'' in \emph{IEEE International
  Symposium on Information Theory (ISIT)}, Athens, Greece, Jul. 2024, pp.
  3444--3449.

\bibitem{morozov2024shannon}
------, ``Shannon capacity of channels with {Markov} insertions, deletions and
  substitutions,'' \emph{arXiv preprint arXiv:2401.16063}, 2024.

\bibitem{gallager2000sequential}
R.~G. Gallager, \emph{Sequential decoding for binary channels with noise and
  synchronization errors}.\hskip 1em plus 0.5em minus 0.4em\relax British
  Library, Reports \& Microfilms, 2000.

\bibitem{zigangirov1969sequential}
K.~Zigangirov, ``Sequential decoding for a binary channel with drop-outs and
  insertions,'' \emph{Problemy Peredachi Informatsii}, vol.~5, no.~2, pp.
  23--30, 1969.

\bibitem{mitzenmacher2006simple}
M.~Mitzenmacher and E.~Drinea, ``A simple lower bound for the capacity of the
  deletion channel,'' \emph{IEEE Transactions on Information Theory}, vol.~52,
  no.~10, pp. 4657--4660, Oct. 2006.

\bibitem{kirsch2009directly}
A.~Kirsch and E.~Drinea, ``Directly lower bounding the information capacity for
  channels with iid deletions and duplications,'' \emph{IEEE Transactions on
  Information Theory}, vol.~56, no.~1, pp. 86--102, Jan. 2009.

\bibitem{diggavi2007capacity}
S.~Diggavi, M.~Mitzenmacher, and H.~D. Pfister, ``Capacity upper bounds for the
  deletion channel,'' in \emph{IEEE International Symposium on Information
  Theory}, Nice, France, Jun. 2007, pp. 1716--1720.

\bibitem{fertonani2010novel}
D.~Fertonani and T.~M. Duman, ``Novel bounds on the capacity of the binary
  deletion channel,'' \emph{IEEE Transactions on Information Theory}, vol.~56,
  no.~6, pp. 2753--2765, Jun. 2010.

\bibitem{rahmati2013upper}
M.~Rahmati and T.~M. Duman, ``An upper bound on the capacity of non-binary
  deletion channels,'' in \emph{IEEE International Symposium on Information
  Theory}, Istanbul, Turkey, Jul. 2013, pp. 2940--2944.

\bibitem{fertonani2010bounds}
D.~Fertonani, T.~M. Duman, and M.~F. Erden, ``Bounds on the capacity of
  channels with insertions, deletions and substitutions,'' \emph{IEEE
  Transactions on Communications}, vol.~59, no.~1, pp. 2--6, Jan. 2010.

\bibitem{10.1145/3281275}
M.~Cheraghchi, ``Capacity upper bounds for deletion-type channels,''
  \emph{Journal of the ACM (JACM)}, vol.~66, no.~2, pp. 1--79, Mar. 2019.

\bibitem{cheraghchi2020overview}
M.~Cheraghchi and J.~Ribeiro, ``An overview of capacity results for
  synchronization channels,'' \emph{IEEE Transactions on Information Theory},
  vol.~67, no.~6, pp. 3207--3232, Jun. 2020.

\bibitem{5513746}
A.~Kalai, M.~Mitzenmacher, and M.~Sudan, ``Tight asymptotic bounds for the
  deletion channel with small deletion probabilities,'' in \emph{IEEE
  International Symposium on Information Theory}, Jun. 2010, pp. 997--1001.

\bibitem{kanoria2010deletion}
Y.~Kanoria and A.~Montanari, ``On the deletion channel with small deletion
  probability,'' in \emph{IEEE International Symposium on Information Theory},
  Austin, TX, USA, Jun. 2010, pp. 1002--1006.

\bibitem{kanoria2013optimal}
------, ``Optimal coding for the binary deletion channel with small deletion
  probability,'' \emph{IEEE Transactions on Information Theory}, vol.~59,
  no.~10, pp. 6192--6219, Oct. 2013.

\bibitem{6457365}
M.~Ramezani and M.~Ardakani, ``On the capacity of duplication channels,''
  \emph{IEEE Transactions on Communications}, vol.~61, no.~3, pp. 1020--1027,
  Mar. 2013.

\bibitem{4418490}
M.~Mitzenmacher, ``Capacity bounds for sticky channels,'' \emph{IEEE
  Transactions on Information Theory}, vol.~54, no.~1, pp. 72--77, Jan. 2008.

\bibitem{tegin2024capacity}
B.~Tegin and T.~M. Duman, ``Capacity approximations for insertion channels with
  small insertion probabilities,'' \emph{arXiv preprint arXiv:2411.14771},
  2024.

\end{thebibliography}

\end{document}